\documentstyle{aipproc}

\begin{document}
\title{Gamma and Cosmic-Ray Tests \\
of Special Relativity}

\author{Luis Gonzalez-Mestres}
\address{CNRS-IN2P3, B.P. 110, 74941 Annecy-le-Vieux Cedex, France\\
E-mail: gonzalez@lappa.in2p3.fr , telephone and fax: 33 1 45830720}

\maketitle

\begin{abstract}
Lorentz symmetry violation (LSV) at Planck scale 
can be tested (see e.g. physics/0003080) through 
ultra-high energy cosmic rays (UHECR). In a deformed Lorentz symmetry 
(DLS) pattern where 
the effective LSV 
parameter varies like the square
of the momentum scale (quadratically deformed relativistic kinematics,
QDRK), a $\approx ~10^{-6}$ LSV at Planck scale
would be
enough to produce observable effects on the properties of
cosmic rays at the $\approx ~10^{20}~eV$ scale: absence of GZK cutoff,
stability of unstable particles, lower interaction rates, kinematical
failure
of any parton model and of standard formulae for Lorentz contraction
and time dilation... Its phenomenological implications are compatible 
with existing data. If the effective LSV parameter is taken to vary linearly 
with the momentum scale (linearly deformed relativistic kinematics, 
LDRK), a LSV at Planck scale larger than $\approx ~10^{-7}$ 
seems to lead to contradictions with data above $\approx ~TeV$ 
energies. 
Consequences are important for high-energy gamma-ray 
experiments, as well as for high-energy cosmic rays and gravitational waves.
\end{abstract}

\section*{QDRK}

It
has been pointed out \cite{gon00} that Lorentz symmetry violation (LSV) at
Planck scale can  
play a crucial role in astrophysical processes at very
high energy. Ultra-high energy (UHE) cosmic rays \cite{NAG}
provide a laboratory to test this prediction. 
A typical pattern 
is provided by models with an absolute local rest frame and
a fundamental length scale $a$ where new physics
is expected to occur \cite{gon97a}.
They naturally lead to a quadratically deformed
relativistic kinematics (QDRK) of the form \cite{gon97b}:

\equation
E~=~~(2\pi )^{-1}~h~c~a^{-1}~e~(k~a)
\endequation
\noindent
where $h$ is the Planck constant, $c$ the speed of light, $k$ the wave vector,
and
$[e~(k~a)]^2$ is a convex
function of $(k~a)^2$ obtained from vacuum dynamics.
Expanding equation (1) for $k~a~\ll ~1$ , we can write
\cite{gon97b}:
\begin{eqnarray}
e~(k~a) & \simeq & [(k~a)^2~-~\alpha ~(k~a)^4~
+~(2\pi ~a)^2~h^{-2}~m^2~c^2]^{1/2}
\end{eqnarray}
\noindent
$\alpha $ being a model-dependent constant, in the range $0.1~-~0.01$ for
full-strength violation of Lorentz symmetry at the fundamental length scale,
and {\it m} the mass of the particle. For momentum $p~\gg ~mc$ , we get:
\begin{eqnarray}
E & \simeq & p~c~+~m^2~c^3~(2~p)^{-1}~
-~p~c~\alpha ~(k~a)^2/2~~~~~
\end{eqnarray}
It is assumed that the earth moves slowly with
respect to the absolute rest frame.
The "deformation" approximated by
$\Delta ~E~=~-~p~c~\alpha ~(k~a)^2/2$ in the right-hand
side of (3) implies a Lorentz symmetry violation in the ratio $E~p^{-1}$
varying like $\Gamma ~(k)~\simeq ~\Gamma _0~k^2$ where $\Gamma _0~
~=~-~\alpha ~a^2/2$ . If $c$ is a universal parameter for all
particles, the QDRK defined by (1) - (3) preserves Lorentz symmetry
in the limit $k~\rightarrow ~0$, contrary to the standard
$TH\epsilon \mu $ model \cite{will} . 

At energies above
$E_{trans}~
\approx ~\pi ^{-1/2}~ h^{1/2}~(2~\alpha )^{-1/4}~a^{-1/2}~m^{1/2}~c^{3/2}$,
the deformation $\Delta ~E$
dominates over
the mass term $m^2~c^3~(2~p)^{-1}$ in (3) and modifies all
kinematical balances:
physics gets closer to Planck scale than
to electroweak scale 
and ultra-high energy cosmic rays (UHECR) 
become an efficient probe of Planck-scale physics. The parton model (in any 
version) does no longer hold, and similarly for standard formulae on Lorentz 
contraction and time dilation \cite{gon97c}
Because of the negative value of $\Delta ~E$ \cite{gon97d} , it costs
more and more energy, as $E$ increases,
to split the incoming logitudinal momentum in the laboratory rest frame.
As the ratio $m^2~c^3~(2~p~\Delta ~E)^{-1}$ varies like $\sim ~E^{-4}$ ,
the transition at $E_{trans}$ is very sharp. QDRK can
lead \cite{gon00} to important observable phenomena. In particular:

- For $\alpha ~a^2~>~10^{-72}~cm^2$ ,
and assuming universal values of $\alpha $ and $c$ ,
there is no GZK
cutoff 
for the particles under
consideration \cite{gon97a} . In astrophysical processes at very
high energy,
similar mechanisms can inhibit \cite{gon97d} radiation under
external forces (e.g. synchrotron-like, where the interactions occur with
virtual photons), 
photodisintegration of nuclei, momentum loss trough
collisions (e.g. with a photon wind in reverse shocks),
production of lower-energy secondaries...

- Unstable particles with at
least two stable particles in the final states
of all their decay channels become stable at very
high energy \cite{gon97a}. Above $E_{trans}$, the lifetimes of all
unstable particles (e.g. the $\pi ^0$ in
cascades) become much longer than predicted
by relativistic kinematics. 
The neutron or even the $\Delta ^{++}$ can be candidates for the
primaries of the highest-energy cosmic ray
events. If $c$ and $\alpha $ are not exactly universal,
many different scenarios are possible 
\cite{gon97d} .

- The allowed final-state
phase space of two-body collisions is strongly
reduced at very high energy,  
\cite{gon97e} , 
with a sharp fall of partial and total cross-sections
for cosmic-ray energies above
$E_{lim} ~\approx ~(2~\pi )^{-2/3}~(E_T~a^{-2}~ \alpha ^{-1}~h^2~c^2)^{1/3}$,
where $E_T$ is the target energy.
Using the
previous figures for LSV parameters, above some
energy $E_{lim}$ between 10$^{22}$ and $10^{24}$ $eV$ a cosmic
ray will not deposit most of its energy in the atmosphere
and can possibly fake an exotic event with much less energy \cite{gon97e} .

Requiring simultaneously the absence of GZK cutoff in the region
$E~\approx ~
10^{20}~eV$~, and that cosmic rays with
$E$ below $\approx ~3.10^{20}~eV$ deposit most of their energy in the
atmosphere, leads to the constraint \cite{gon97e} :
$10^{-72}~cm^2~<~\alpha ~a^2~<~
10^{-61}~cm^2$~, equivalent to $10^{-20}~<~\alpha ~<~10^{-9}$ for
$a~\approx 10^{-26}~cm$~ ($\approx~10^{21}~GeV$ scale).
Assuming full-strength
LSV forces $a$ to be in the range $10^{-36}~cm~<~a~<~
10^{-30}~cm$ , but a $\approx 10^{-6}$ LSV at Planck scale
can still explain the data. Thus, the simplest
version of QDRK naturally fits
with the expected potential
role of Planck-scale dynamics. 

\section*{LDRK}

LDRK, linearly deformed relativistic kinematics, was discarded in our 1997 
and subsequent papers for phenomenological reasons \cite{gon00} , but has 
nevertheless been proposed by several authors (see e.g. \cite{amel}), 
for cosmic-ray and gravitational-wave phenomenology. The 
fonction $e~(k~a)$ is then a
function of $k~a$ and, for $k~a~\ll ~1$ :
\begin{eqnarray}
e~(k~a) ~ \simeq ~ [(k~a)^2~-~\beta ~(k~a)^3~
+~(2\pi ~a)^2~h^{-2}~m^2~c^2]^{1/2}
\end{eqnarray}
\noindent
$\beta $ being a model-dependent constant.
For momentum $p~\gg ~mc$ :
\begin{eqnarray}
E ~ \simeq ~ p~c~+~m^2~c^3~(2~p)^{-1}~
-~p~c~\beta ~(k~a)/2~~~~~
\end{eqnarray}
\noindent
the deformation $\Delta~E~=~-~p~c~\beta ~(k~a)/2$ being now driven by an
effective parameter proportional to momentum.
LDRK can be generated by
introducing a background
gravitational field in the propagation equations of free particles \cite{el1}~.
If existing bounds on LSV from
nuclear magnetic resonance experiments are to be intepreted as setting a
bound of $\approx 10^{-21}$ on relative LSV at the momentum scale
$p~\sim ~100~MeV$ , this implies $\beta ~a~<~10^{-34}~cm$ . But LDRK leads to
inconsistencies with cosmic-ray experiments unless $\beta ~a$ is much
smaller. Concepts and
formulae presented for QDRK can be readily extended to LDRK, and we get now: 
\begin{eqnarray}
E_{trans}~
\approx ~\pi ^{-1/3}~ h^{1/3}~(2~\beta )^{-1/3}~a^{-1/3}~m^{2/3}~c^{5/3} \\
E_{lim} ~\approx ~(2~\pi )^{-1/2}~(E_T~a^{-1} \beta ^{-1}~h~c)^{1/2}~~~~~~
\end{eqnarray}
For a high-energy photon, LDRK is usually parameterized \cite{el1} as:
\begin{eqnarray}
E ~\simeq ~p~c~-~p~c~\beta ~(k~a)/2~=~p~c~-~p^2~M^{-1}
\end{eqnarray} 
where $M$ is an effective mass scale. Tests of this model through
gamma-ray bursts, measuring the delays in the arrival time of photons of
different energies, have been considered in \cite{norris} for 
the Gamma-ray Large Area Space Telescope (GLAST), and 
more generally in \cite{el1} . 
But, from the same considerations developed in our 1997-99
papers taking QDRK as an exemple, stringent bounds on LDRK can be
derived.
Assume that LDRK applies only to photons, and not to charged particles, so
that at high energy we can write for a charged particle, $ch$ ,
the dispersion
relation:

\begin{eqnarray}
E_{ch} & \simeq & p_{ch}~c~+~m_{ch}^2~c^3~(2~p_{ch})^{-1}
\end{eqnarray}
\noindent
where the $ch$ subscript stands for the charged particle under consideration.
Then, it can be readily checked that the decay $ch~\rightarrow ~ch~+~\gamma$
would be allowed for $p$ above $\simeq ~(2~m_{ch}^2~M~c^3)^{1/3}$ , i.e:

- for an electron, above $E ~\approx ~2~TeV$ if $M~=10^{16}~GeV$ ,
and above $\approx ~20~TeV$ if $M~=10^{19}~GeV$ ;

- for a muon or charged pion, above $E ~\approx ~80~TeV$ if $M~=10^{16}~GeV$ ,
and above $\approx ~800~TeV$ if $M~=10^{19}~GeV$ ;

- for a proton, above $E ~\approx ~240~TeV$ if $M~=10^{16}~GeV$ ,
and above $\approx ~2.4~PeV$ if $M~=10^{19}~GeV$ ;

- for a $\tau $ lepton, above $E ~\approx ~400~TeV$ if $M~=10^{16}~GeV$ ,
and above $\approx ~4~PeV$ if $M~=10^{19}~GeV$ ; 

\noindent
so that none of these particles would be oberved above such energies,
apart from very short paths.
Such decays seem to be in contradiction with cosmic ray data, but
avoiding them forces the charged particles to have the same kind of
propagators as the photon, with the same effective value of $M$ up to small
differences. Similar conditions are readily derived for all
"elementary" particles, leading for all of them, up to small devations,
to a LDRK given by the universal dispersion relation:
\begin{eqnarray}
E ~\simeq ~p~c~+~m^2~c^3~(2~p)^{-1}~-~p^2~M^{-1}
\end{eqnarray}

For instance, $\pi^0$ production would otherwise be
inhibited. But if, as it seems compulsory, the $\pi^0$ kinematics
follows a similar law, then the decay time for $\pi^0 ~\rightarrow ~\gamma
~\gamma $ will become much longer than predicted by special relativity at
energies above  $\approx ~50~TeV$ if $M~=10^{16}~GeV$
and $\approx ~500~TeV$ if $M~=10^{19}~GeV$ . Again, this seems to
be in contradiction
with cosmic-ray data. Requiring that the $\pi^0$ lifetime agrees
with special relativity at $E ~\approx ~10^{17}~eV$ would force $M$
to be above $\approx ~10^{26}~GeV$ , far away from the values to be
tested at GLAST. Another bound is obtained from the
condition that there are $3.10^{20}~eV$ cosmic-ray events. Setting
$E_{lim}$ to this value, and taking oxygen to be the
target, yields $M~\approx ~
3.10^{21}GeV$ . As it
appears very difficult to make LDRK , with $M$
reasonably close to Planck scale, compatible with experimental data,
it seems necessary to reconsider the models to be tested at GLAST
and in gravitational-wave experiments.
~

\end{document}